\documentclass{IEEEtran}
\usepackage{cite}
\usepackage{amsmath,amssymb,amsfonts}
\usepackage{algorithmic}
\usepackage{graphicx}
\usepackage{textcomp}
\usepackage{enumerate}
\usepackage{amsmath}
\usepackage{amsfonts}

\def\BibTeX{{\rm B\kern-.05em{\sc i\kern-.025em b}\kern-.08em
    T\kern-.1667em\lower.7ex\hbox{E}\kern-.125emX}}

\begin{document}

\title{State Estimation of Wireless Sensor Networks in the Presence of Data Packet Drops and Non-Gaussian Noise} 
\author{Jiacheng He, Gang Wang, Xuemei Mao, Song Gao, Bei Peng}

\maketitle

\begin{abstract}
Distributed Kalman filter approaches based on the maximum correntropy criterion have recently demonstrated superior state estimation performance to that of conventional distributed Kalman filters for wireless sensor networks in the presence of non-Gaussian impulsive noise. However, these algorithms currently fail to take account of data packet drops. The present work addresses this issue by proposing a distributed maximum correntropy Kalman filter that accounts for data packet drops (i.e., the DMCKF-DPD algorithm). The effectiveness and feasibility of the algorithm are verified by simulations conducted in a wireless sensor network with intermittent observations due to data packet drops under a non-Gaussian noise environment. Moreover, the computational complexity of the DMCKF-DPD algorithm is demonstrated to be moderate compared with that of a conventional distributed Kalman filter, and we provide a sufficient condition to ensure the convergence of the proposed algorithm.
\end{abstract}

\IEEEpeerreviewmaketitle

\section{Introduction}
State estimation in wireless sensor networks (WSNs) has been a subject of considerable interest recently\cite{YANG2019156,feng2023distributed,zheng2018consensus}, and it has been applied widely, such as in static or dynamic target positioning and tracking\cite{8769946,wang2018maximum}, indoor positioning\cite{9031316,he2023generalized11}, vehicle navigation\cite{8613889,wang2018complex}, and smart grids\cite{8187698}. However, a number of challenges affecting the state estimation performance in WSNs remain to be fully addressed, such as those due to data packet drops and non-Gaussian noise \cite{hongwei2015comparison}, which are both very common in practical applications of automatic control and target tracking.

The application of Kalman filters to the state estimation problem in WSNs with intermittent observations due to data packet drops has undergone significant development recently\cite{1272646,1333199,KARIMI201949,9123558,8920146}. The development of distributed Kalman filter (DKF) approaches has demonstrated particularly great progress in this regard\cite{YANG2019156,wang2014information,7417484,8409298,alam2014distributed,6374678,he2023maximum}. In a DKF approach, each sensor in a WSN computes local state estimation based on the information obtained from its own observations and the observations sent from its neighbors. In addition, algorithms based on the consensus approach have been proposed to address random link failure in WSNs due to data packet drops\cite{7417484}. This issue has also been addressed by developing a novel state estimation algorithm based on a distributed agent dynamical system\cite{alam2014distributed}. Zhou \emph{et al.}\cite{8409298} considered the existence of data packet drops when designing a DKF for WSNs, and the performance of a stationary DKF and a Kalman consensus filter (KCF) were also evaluated. However, all of the above-discussed algorithms have accounted for the impact of data packet drops in the state estimation process under a Gaussian noise assumption.

A number of studies have focused on conducting state estimation in WSNs under a non-Gaussian noise environment \cite{fan2022background,he2023generalized,zhong2023pseudolinear}. For example, cost functions based on information theoretic learning (ITL) have been proposed\cite{wang2018complex,liu2007correntropy,wang2018switching,chen2017maximum,8424469,WANG20178659,zhao2017projected}. In addition, approaches based on the maximum correntropy (MC) criterion have demonstrated superior state estimation performance under conditions of impulsive noise\cite{liu2007correntropy,wang2018switching,8718566,wang2018switching}. Moreover, a proposed DFK algorithm based on MC (i.e., the DMCKF algorithm) has been demonstrated to significantly improve the state estimation performance under non-Gaussian noise conditions\cite{wang2019distributed,he2022mixture,ye2009efficient}. However, all of the above-discussed algorithms fail to take account of data packet drops in the state estimation process of WSNs.

The present study addresses the above-discussed issues by proposing a novel DMCKF algorithm accounting for data packet drops (i.e., the DMCKF-DPD algorithm) for conducting state estimation in WSNs. In contrast to conventional DMCKF algorithms, the proposed algorithm is based on a fixed-point iterative algorithm to update the posterior estimates of the WSN node states. The effectiveness and feasibility of the algorithm are verified by simulations conducted in a WSN with intermittent observations due to data packet drops under a non-Gaussian noise environment. Moreover, the computational complexity of the DMCKF-DPD algorithm is demonstrated to be moderate compared with that of a conventional stationary DKF\cite{8409298}, and we provide a sufficient condition to ensure the convergence of the proposed algorithm.

The remainder of this paper is organized as follows. In Section \ref{section:Preliminaries}, the MC criterion and state-space model are briefly reviewed. In Section \ref{section:derivation}, the derivation, computational complexity, and convergence issue of the DMCKF-DPD algorithm are presented. The simulation examples are provided in Section \ref{section:simulations}, and, finally, the conclusion is given in Section \ref{section:Conclusion}.

\section{Preliminaries} \label{section:Preliminaries}
\subsection{Correntropy}
The concept of correntropy, first proposed by Principe \emph{et al.}\cite{liu2007correntropy}, is a very practical method for evaluating the generalized similarity between random variables ${X,Y \in R}$ with the same dimensions. Here, correntropy is defined as
\begin{equation}
\begin{split}
\operatorname{V} \left( {X,Y} \right) = \operatorname{E} \left[ {\kappa \left( {X,Y} \right)} \right] = \int {\kappa \left( {x,y} \right)} \operatorname{d} {\operatorname{F} _{XY}}\left( {x,y} \right),
\end{split}
\end{equation}
where ${{\text{E[}}{\text{.]}}}$ is the expectation operator, ${\operatorname{V} (.)}$ represents the information potential, ${{\operatorname{F} _{XY}}(x,y)}$ represents the probability distribution function (PDF) with respect to variables ${X}$ and ${Y}$, and ${\kappa \left( { \cdot {\rm{ }},{\rm{ }} \cdot } \right)}$ is the shift-invariant Mercer kernel. Here, we employ the Gaussian kernel, which is given as
\begin{equation}\label{equ:epxeeg}
\begin{split}
\kappa \left( {x,y} \right) = {\operatorname{G} _\sigma }\left( e \right) = \exp \left( { - \frac{{{e^2}}}{{2{\sigma ^2}}}} \right),
\end{split}
\end{equation}
where ${e = x - y}$ represents the error between elements ${x}$ and ${y}$, and ${\sigma  > 0}$ represents the kernel bandwidth (or kernel size) of  the Gaussian kernel function.

However, only a limited amount of data related to the variables ${X}$ and ${Y}$ can be obtained in realistic scenarios, and the PDF ${{\operatorname{F} _{XY}}(x,y)}$ is usually unknown. Under these conditions, a sample estimator can be used to calculate the correntropy as follows:
\begin{equation}
\begin{split}
\hat V\left( {X,Y} \right) = \frac{1}{N}\sum\limits_{i = 1}^N {{\operatorname{G} _\sigma }} \left( {{e^i}} \right),
\end{split}
\end{equation}
where
\begin{equation}
\begin{split}
{e^i} = {x^i} - {y^i},\left( {{x^i},{y^i} \in \left\{ {{x^i},{y^i}} \right\}_{i = 1}^N} \right),
\end{split}
\end{equation}
and ${N}$ samples are employed to define ${{\operatorname{F} _{XY}}(x,y)}$. 

Applying a Taylor expansion to the Gaussian kernel function yields the following:
\begin{equation}\label{equ:n2yjxkke}
\begin{split}
\hat V\left( {X,Y} \right) = \sum\limits_{n = 0}^\infty  {\frac{{{{\left( { - 1} \right)}^n}}}{{{2^n}{\sigma ^{2n}}n!}}\operatorname{E} \left[ {{{\left( {X - Y} \right)}^{2n}}} \right]} .
\end{split}
\end{equation}
We can infer from (\ref{equ:n2yjxkke}) that the correntropy is the weighted sum of all even-order moments of relation ${X - Y}$. Compared with other similarity measurement schemes, such as the mean-square error (MSE) criterion, correntropy contains all even-order moments, and is therefore useful for nonlinear and signal processing applications in non-Gaussian noise environments.
\subsection{State-space model}
The state-space model of the DKF over the nodes of a WSN under conditions of data packet drops is the first problem that must be solved to estimate the state of a target node. Here, we adopt a powerful state-space model \cite{8409298}, which is described as follows. 

Consider a WSN with ${N}$ sensors. The states and observations are as follows:
\begin{equation}
\begin{split}
\begin{gathered}
  {{\boldsymbol{x}}_k} = {{\boldsymbol{A}}_k}{{\boldsymbol{x}}_{k - 1}} + {{\boldsymbol{q}}_k}, \hfill \\
  {\boldsymbol{y}}_k^i = {{\boldsymbol{C}}^i}{\boldsymbol{x}}_k^i + {\boldsymbol{v}}_k^i,{\text{  }}i = 1,2, \cdots ,N,{\text{and }}i \in \Omega , \hfill \\ 
\end{gathered}  
\end{split}
\end{equation}
where ${{{\boldsymbol{x}}_k},{{\boldsymbol{x}}_{k - 1}} \in {\mathbb{R}^{n \times 1}}}$ represent the states of the dynamical system at instants ${k}$ and ${k - 1}$, ${\Omega }$ represents the set of all sensors in the network, ${{\boldsymbol{y}}_k^i \in {\mathbb{R}^{{m_i} \times 1}}({m_i} = \operatorname{rank} ({{\boldsymbol{C}}_i})\forall i \in \Omega )}$ denotes the observations obtained by node ${i}$ at instant ${k}$, ${{{\boldsymbol{A}}_k}}$ and ${{{\boldsymbol{C}}^i}}$ represent the state transition matrix and observation matrix of the system, respectively, and ${{{\boldsymbol{q}}_k}}$ and ${{\boldsymbol{v}}_k^i}$ are the mutually uncorrelated process noise and measurement noise, respectively, with means of zero and covariances respectively given as follows:
\begin{equation}
\begin{split}
{\text{E}}\left[ {{{\boldsymbol{w}}_k}{\boldsymbol{w}}_k^{\text{T}}} \right] = {\boldsymbol{P}},{\text{ E}}\left[ {{\boldsymbol{v}}_k^i{{({\boldsymbol{v}}_k^i)}^{\text{T}}}} \right] = {{\boldsymbol{R}}^i}.
\end{split}
\end{equation}

We then define ${{\Omega _i} \subset \Omega }$ to represent the set of all neighboring sensors of the ${i}$th sensor. Therefore, set ${{\mho _i} = {\Omega _i} \cup \{ i\}}$ contains node ${i}$ itself and all its neighboring nodes. According to the properties of a WSN, the process whereby node ${i}$ receives the observations from its neighboring node ${j}$ at instant ${k}$ can be modeled as follows:
\begin{equation}
\begin{split}\label{equ:iosjoosjkj}
{\boldsymbol{s}}_k^{i,j} = \gamma _k^{i,j}{\boldsymbol{y}}_k^j,{\text{  }}i \in \Omega ,{\text{  }}j \in {\mho _i}.
\end{split}
\end{equation}
Here, the term ${\gamma _k^{i,j}}$ is used to indicate the existence of data packet drops over the WSN. We employ the Bernoulli model for modeling the data packet loss process and assume that the process has independent and identically distributed properties for ${\gamma _k^{i,j}}$ with the probability
\begin{equation}
\begin{split}
\operatorname{P} \left\{ {\gamma _k^{i;j} = 1} \right\} = p_k^{i;j} > 0{\text{  }}\forall k \geqslant 0,
\end{split}
\end{equation}
and 
\begin{equation}
\begin{split}
\begin{gathered}
  \operatorname{E} \left[ {\gamma _k^{i;j} = 1} \right] = p_k^{i;j}, \hfill \\
  {\left( {\mu _k^{i;j}} \right)^2} = \operatorname{E} \left[ {{{\left( {\gamma _k^{i;j} - p_k^{i;j}} \right)}^2}} \right] = p_k^{i;j} - {\left( {p_k^{i;j}} \right)^2}. \hfill \\ 
\end{gathered}   
\end{split}
\end{equation}
We assume that variables ${\gamma _k^{i,j}}$ and ${\gamma _k^{j,i}}$ are mutually independent (i.e., ${\forall i \ne j}$), but the condition ${{p_{i,j}} = {p_{j,i}}}$ is allowed, and ${\gamma _k^{i,j}}$ is independent of the process noise, measurement noise, and initial state of the dynamical system. According to (\ref{equ:iosjoosjkj}), if sensor ${i}$ receives the observations from its neighboring node ${j}$ at instant ${k}$, then ${\gamma _k^{i,j} = 1}$ and ${\gamma _k^{i,j} = 0}$ when all the components of ${{{\boldsymbol{y}}_k^j}}$ are lost. We assume that node ${i}$ can receive all observations from itself at any time, so ${\gamma _k^{i,j} \equiv 1}$. 

Based on the above analysis, the various factors of a DKF model that must take data packet drops into account can be expressed as
\begin{equation}
\begin{split}
\left\{ \begin{gathered}
  {\boldsymbol{y}}_k^{{\mho _i}} = {\text{vec}}{\left\{ {{\boldsymbol{y}}_k^j} \right\}_{j \in {\mho _i}}}{\text{,}} \hfill \\
  {\boldsymbol{v}}_k^{{\mho _i}}{\text{ }} = {\text{vec}}{\left\{ {{\boldsymbol{v}}_k^j} \right\}_{j \in {\mho _i}}}, \hfill \\
  {\boldsymbol{s}}_k^{{\mho _i}} = {\text{vec}}{\left\{ {{\boldsymbol{s}}_k^j} \right\}_{j \in {\mho _i}}}, \hfill \\
  {{\boldsymbol{C}}^{{\mho _i}}} = {\text{col}}{\left\{ {{{\boldsymbol{C}}^j}} \right\}_{j \in {\mho _i}}}, \hfill \\
  {{\boldsymbol{R}}^{{\mho _i}}} = \operatorname{diag} {\left\{ {{{\boldsymbol{R}}^j}} \right\}_{j \in {\mho _i}}}, \hfill \\
  {\boldsymbol{D}}_{\gamma ;k}^{{\mho _i}} = \operatorname{diag} {\left\{ {\gamma _k^{i,j}{{\boldsymbol{I}}_{{m_j}}}} \right\}_{j \in {\mho _i}}}. \hfill \\ 
\end{gathered}  \right.
\end{split}
\end{equation}
Here, ${{\text{vec}}\left\{  \cdot  \right\}}$, ${{\text{col}}\left\{  \cdot  \right\}}$, and ${\operatorname{diag} \left\{  \cdot  \right\}}$ denote the vectorization operation, the columnization operation, and the (block) diagonalization operation, respectively, and  ${{{\boldsymbol{I}}_{{m_j}}}(j \in {\mho _i})}$ represents an identity matrix of dimension ${{m_j} \times {m_j}}$. According to the above discussion, the state-space model of the ${i}$th sensor in the case of data packet drops can be written in a compact form as
\begin{equation}\label{equ:kiqj1jkisaa}
\begin{split}
{\boldsymbol{x}}_k^i = {\boldsymbol{Ax}}_{k - 1}^i + {\boldsymbol{q}}_k^i,
\end{split}
\end{equation}
\begin{equation}\label{equ:aa}
\begin{split}
{\boldsymbol{y}}_k^{{\mho _i}} = {{\boldsymbol{C}}^{{\mho _i}}}{\boldsymbol{x}}_k^i + {\boldsymbol{v}}_k^{{\mho _i}},
\end{split}
\end{equation}
\begin{equation}\label{equ:kiyikd}
\begin{split}
{\boldsymbol{s}}_k^{{\mho _i}} = {\boldsymbol{D}}_k^{\gamma ;{\mho _i}}{\boldsymbol{y}}_k^{{\mho _i}}.
\end{split}
\end{equation}
In addition, the covariance of ${{\boldsymbol{v}}_k^{{\mho _i}}}$ is 
\begin{equation}
\begin{split}
{{\boldsymbol{R}}^{{\mho _i}}} = \operatorname{diag} {\left\{ {{{\boldsymbol{R}}^j}} \right\}_{j \in {\mho _i}}}.
\end{split}
\end{equation}

\section{Proposed DMCKF-DPD algorithm}\label{section:derivation}
\subsection{Algorithm derivation}
We first derive the stationary DMCKF in the case of data packet drops based on the model given by (\ref{equ:kiqj1jkisaa})-(\ref{equ:kiyikd}). The state prediction error of the algorithm can be written as
\begin{equation}
\begin{split}
{\boldsymbol{\varepsilon }}_{k|k - 1}^i = {\boldsymbol{x}}_k^i - {\boldsymbol{\hat x}}_{k|k - 1}^i.
\end{split}
\end{equation}
Combining this with the above state-space model yields the following augmented model:
\begin{equation}\label{equ:kigjkixkio}
\begin{split}
\left[ {\begin{array}{*{20}{c}}
  {{\boldsymbol{\hat x}}_{k|k - 1}^i} \\ 
  {{\boldsymbol{s}}_k^{{\mho _i}}} 
\end{array}} \right] = \left[ {\begin{array}{*{20}{c}}
  {{{\boldsymbol{I}}_n}} \\ 
  {{\boldsymbol{D}}_{\gamma ;k}^{{\mho _i}}{{\boldsymbol{C}}^{{\mho _i}}}} 
\end{array}} \right]{\boldsymbol{x}}_k^i{\text{ + }}{\boldsymbol{g}}_k^i,
\end{split}
\end{equation}
where ${{{{\boldsymbol{I}}_n}}}$ represents an ${n \times n}$ identity matrix and ${{\boldsymbol{g}}_k^i}$ represents the augmented noise vector containing the state and measurement errors of the dynamical system, which is defined as 
\begin{equation}
\begin{split}
{\boldsymbol{g}}_k^i = \left[ {\begin{array}{*{20}{c}}
  { - {\boldsymbol{\varepsilon }}_{k|k - 1}^i} \\ 
  {{\boldsymbol{D}}_{\gamma ;k}^{{\mho _i}}{\boldsymbol{v}}_k^{{\mho _i}}} 
\end{array}} \right].
\end{split}
\end{equation}
Assuming that the covariance matrix of the augmented vector ${{\text{E}}[{\boldsymbol{g}}_k^i{({\boldsymbol{g}}_k^i)^{\text{T}}}]}$ is positive definite yields the following:
\begin{equation}\label{equ:tkikbkikb}
\begin{split}
\begin{gathered}
  {\text{E}}\left[ {{\boldsymbol{g}}_k^i{{\left( {{\boldsymbol{g}}_k^i} \right)}^{\text{T}}}} \right] = \left[ {\begin{array}{*{20}{c}}
  {{\boldsymbol{P}}_{k|k - 1}^i}&0 \\ 
  0&{{\boldsymbol{D}}_{p;k}^{{\mho _i}}{{\boldsymbol{R}}^{{\mho _i}}}{\boldsymbol{D}}_{p;k}^{{\mho _i}}} 
\end{array}} \right] \hfill \\
   = \left[ {\begin{array}{*{20}{c}}
  {{\boldsymbol{B}}_{P\left( {k|k - 1} \right)}^i{{\left( {{\boldsymbol{B}}_{P\left( {k|k - 1} \right)}^i} \right)}^{\text{T}}}}&0 \\ 
  0&{{\boldsymbol{B}}_{R;k}^i{{\left( {{\boldsymbol{B}}_{R;k}^i} \right)}^{\text{T}}}} 
\end{array}} \right] \hfill \\
   = {\boldsymbol{B}}_k^i{\left( {{\boldsymbol{B}}_k^i} \right)^{\text{T}}}, \hfill \\ 
\end{gathered}    
\end{split}
\end{equation}
with 
\begin{equation}
\begin{split}
{\text{E}}\left[ {{\boldsymbol{D}}_{\gamma ;k}^{{\mho _i}}} \right] = \operatorname{diag} {\left\{ {p_k^{i;j}{{\boldsymbol{I}}_{{m_j}}}} \right\}_{j \in {\mho _i}}} = {\boldsymbol{D}}_{p;k}^{{\mho _i}}. 
\end{split}
\end{equation}
Here, ${{\boldsymbol{B}}_k^i}$ and ${{({\boldsymbol{B}}_k^i)^{\text{T}}}}$ can be obtained by the Cholesky decomposition of ${{\text{E}}[{\boldsymbol{g}}_k^i{({\boldsymbol{g}}_k^i)^{\text{T}}}]}$, and matrix ${{\boldsymbol{D}}_{p;k}^{{\mho _i}}}$ represents the expectation of ${{{\boldsymbol{D}}_{\gamma ;k}^{{\mho _i}}}}$. Left multiplying each term of (\ref{equ:kigjkixkio}) by $ {{({\boldsymbol{B}}_k^i)^{ - 1}}}$ yields the following:
\begin{equation}
\begin{split}
{\boldsymbol{D}}_k^i = {\boldsymbol{W}}_k^i{\boldsymbol{x}}_k^i + {\boldsymbol{e}}_k^i,
\end{split}
\end{equation}
where 
\begin{equation}\label{equ:jkiociuf}
\begin{split}
\left\{ {\begin{array}{*{20}{l}}
  {{\boldsymbol{D}}_k^i = {{\left( {{\boldsymbol{B}}_k^i} \right)}^{ - 1}}\left[ {\begin{array}{*{20}{c}}
  {{\boldsymbol{\hat x}}_{k|k - 1}^i} \\ 
  {{\boldsymbol{s}}_k^{{\mho _i}}} 
\end{array}} \right],} \\ 
  {{\boldsymbol{W}}_k^i = {{\left( {{\boldsymbol{B}}_k^i} \right)}^{ - 1}}\left[ {\begin{array}{*{20}{c}}
  {\boldsymbol{I}} \\ 
  {{\boldsymbol{D}}_{\gamma ;k}^{{\mho _i}}{{\boldsymbol{C}}^{{\mho _i}}}} 
\end{array}} \right],} \\ 
  {{\boldsymbol{e}}_k^i = {{\left( {{\boldsymbol{B}}_k^i} \right)}^{ - 1}}{\boldsymbol{g}}_k^i.} 
\end{array}} \right.
\end{split}
\end{equation}

According to the above derivation, we propose the following cost function based on the MC criterion:
\begin{equation}
\begin{split}
{J_L}\left( {x_k^i} \right) = \frac{1}{L}\sum\limits_{h = 1}^L {{\operatorname{G} _\sigma }\left( {d_k^{i;h} - {\boldsymbol{w}}_k^{i;h}{\boldsymbol{x}}_k^i} \right)},
\end{split}
\end{equation}
where ${d_k^{i;h}}$ represents the ${h}$th element of ${{\boldsymbol{D}}_k^i}$, ${{\boldsymbol{w}}_k^{i;h}}$ represents the ${h}$th row  of ${{\boldsymbol{W}}_k^i}$, and
\begin{equation}
\begin{split}
L = n + {m_{i;a}},\left( {{{\left( {{m_{i;a}}} \right)}_{i \in \Omega }} = \sum\limits_{j \in {\Omega _i}} {{m_j}} } \right)
\end{split}
\end{equation}
denotes the number of elements of ${{\boldsymbol{D}}_k^i}$. Then, the objective function of the optimal state estimation ${{\boldsymbol{x}}_k^i}$ based on the MC criterion is
\begin{equation}
\begin{split}
{\boldsymbol{\hat x}}_k^i = \mathop {\max }\limits_{{\boldsymbol{x}}_k^i} {J_L}\left( {{\boldsymbol{x}}_k^i} \right) = \mathop {\max }\limits_{{\boldsymbol{x}}_k^i} \sum\limits_{h = 1}^L {{\operatorname{G} _\sigma }} \left( {e_k^{i;h}} \right),
\end{split}
\end{equation}
where ${e_k^{i;h} = d_k^{i;h} - {\boldsymbol{w}}_k^{i;h}{\boldsymbol{x}}_k^i}$ and ${\left( {h = 1,2, \cdots L} \right)}$ is the ${h}$th element of ${e_k^{i}}$. Finally, the optimal state estimation of ${{\boldsymbol{\hat x}}_k^i}$ is achieved by maximizing the information potential ${\operatorname{V} (.)}$. To that end, we set the gradient of the cost function ${{{J_L}({\boldsymbol{x}}_k^i)}}$ regarding ${{{\boldsymbol{x}}_k^i}}$  to zero:
\begin{equation}
\begin{split}
\begin{gathered}
  \frac{{\partial {J_L}\left( {{\boldsymbol{x}}_k^i} \right)}}{{\partial {\boldsymbol{x}}_k^i}} \hfill \\
   = \frac{1}{{{\sigma ^2}}}\sum\limits_{h = 1}^L {{\operatorname{G} _\sigma }\left( {e_k^{i;h}} \right)} {\left( {{\boldsymbol{w}}_k^{i;h}} \right)^{\text{T}}}\left( {d_k^{i;h} - {\boldsymbol{w}}_k^{i;h}{\boldsymbol{x}}_k^i} \right) \hfill \\
   = 0, \hfill \\ 
\end{gathered}    
\end{split}
\end{equation}

and obtaining the optimal state of ${{{\boldsymbol{x}}_k^i}}$ is relatively easy:
\begin{equation}\label{equ:khfidtkkhf}
\begin{split}
\begin{gathered}
  {\boldsymbol{x}}_k^i = {\left[ {\sum\limits_{h = 1}^L {{\operatorname{G} _\sigma }\left( {e_k^{i;h}} \right){{\left( {{\boldsymbol{w}}_k^{i;h}} \right)}^{\text{T}}}} {\boldsymbol{w}}_k^{i;h}} \right]^{ - 1}} \times  \hfill \\
  \sum\limits_{h = 1}^L {{\operatorname{G} _\sigma }\left( {e_k^{i;h}} \right){{\left( {{\boldsymbol{w}}_k^{i;h}} \right)}^{\text{T}}}} d_k^{i;h}. \hfill \\ 
\end{gathered} 
\end{split}
\end{equation}
Because ${e_k^{i;h} = d_k^{i;h} - {\boldsymbol{w}}_k^{i;h}{\boldsymbol{x}}_k^i}$ is a function of ${{\boldsymbol{x}}_k^i}$, the optimal state estimation in (\ref{equ:khfidtkkhf}) is a fixed-point iterative equation of ${{\boldsymbol{x}}_k^i}$, and can be rewritten in the following form:
\begin{equation}\label{equ:dkixkkfddkix}
\begin{split}
{\boldsymbol{x}}_k^i = \operatorname{f} \left( {{\boldsymbol{x}}_k^i} \right),
\end{split}
\end{equation}
where
\begin{equation}
\begin{split}
\begin{gathered}
  \operatorname{f} \left( {{\boldsymbol{x}}_k^i} \right) = {\left[ {\sum\limits_{h = 1}^L {{\operatorname{G} _\sigma }\left( {d_k^{i;h} - {\boldsymbol{w}}_k^{i;h}{\boldsymbol{x}}_k^i} \right)} {{\left( {{\boldsymbol{w}}_k^{i;h}} \right)}^{\text{T}}}{\boldsymbol{w}}_k^{i;h}} \right]^{ - 1}} \hfill \\
  \sum\limits_{h = 1}^L {{\operatorname{G} _\sigma }\left( {d_k^{i;h} - {\boldsymbol{w}}_k^{i;h}{\boldsymbol{x}}_k^i} \right)} {\left( {{\boldsymbol{w}}_k^{i;h}} \right)^{\text{T}}}{\boldsymbol{d}}_k^{i;h}. \hfill \\ 
\end{gathered}  
\end{split}
\end{equation}

According to the above derivation, the iterative equation in (\ref{equ:dkixkkfddkix}) can be written as follows:
\begin{equation}
\begin{split}
{\left( {{\boldsymbol{\hat x}}_k^i} \right)_{t + 1}} = \operatorname{f} \left[ {{{\left( {{\boldsymbol{x}}_k^i} \right)}_t}} \right].
\end{split}
\end{equation}
Here, ${{({\boldsymbol{\hat x}}_k^i)_{t + 1}}}$ represents the result of ${{\boldsymbol{x}}_k^i}$ at the fixed-point iteration ${t+1}$, and (\ref{equ:khfidtkkhf}) can be further written in the form of matrix multiplication:
\begin{equation}\label{equ:kidkiatkwi}
\begin{split}
{\boldsymbol{x}}_k^i = {\left[ {{{\left( {{\boldsymbol{W}}_k^i} \right)}^{\text{T}}}{\boldsymbol{\Lambda }}_k^i{\boldsymbol{W}}_k^i} \right]^{ - 1}}{\left( {{\boldsymbol{W}}_k^i} \right)^{\text{T}}}{\boldsymbol{\Lambda }}_k^i{\boldsymbol{D}}_k^i,
\end{split}
\end{equation}
where 
\begin{equation}\label{equ:jkktjsikesg}
\begin{split}
\begin{gathered}
  {\boldsymbol{\Lambda }}_k^i = \left[ {\begin{array}{*{20}{c}}
  {{\boldsymbol{\Lambda }}_{x;k}^i}&0 \\ 
  0&{{\boldsymbol{\Lambda }}_{y;k}^i} 
\end{array}} \right], \hfill \\
  {\boldsymbol{\Lambda }}_{x;k}^i = {\text{diag}}\left[ {{\operatorname{G} _\sigma }\left( {e_k^{i;1}} \right), \cdots ,{\operatorname{G} _\sigma }\left( {e_k^{i;n}} \right)} \right], \hfill \\
  {\boldsymbol{\Lambda }}_{y;k}^i = {\text{diag}}\left[ {{\operatorname{G} _\sigma }\left( {e_k^{i;n + 1}} \right), \cdots ,{\operatorname{G} _\sigma }\left( {e_k^{i;n + {m_{i;a}}}} \right)} \right]. \hfill \\ 
\end{gathered}     
\end{split}
\end{equation}
According to (\ref{equ:jkiociuf}), (\ref{equ:kidkiatkwi}), and (\ref{equ:jkktjsikesg}), we obtain the following:
\begin{equation}
\begin{split}
\begin{gathered}
  {\left[ {{{\left( {{\boldsymbol{W}}_k^i} \right)}^{\text{T}}}{\boldsymbol{\Lambda }}_k^i{\boldsymbol{W}}_k^i} \right]^{ - 1}} \hfill \\
   = {\left\{ \begin{gathered}
  {\left[ {{{\left( {{\boldsymbol{B}}_{P\left( {k|k - 1} \right)}^i} \right)}^{ - 1}}} \right]^{\text{T}}}{\boldsymbol{\Lambda }}_{x;k}^i{\left( {{\boldsymbol{B}}_{P\left( {k|k - 1} \right)}^i} \right)^{ - 1}} \hfill \\
   + {\left( {{\boldsymbol{D}}_{\gamma ;k}^{{\mho _i}}{{\boldsymbol{C}}^{{\mho _i}}}} \right)^{\text{T}}} \times  \hfill \\
  {\left[ {{{\left( {{\boldsymbol{B}}_{R;k}^i} \right)}^{ - 1}}} \right]^{\text{T}}}{\boldsymbol{\Lambda }}_{y;k}^i{\left( {{\boldsymbol{B}}_{R;k}^i} \right)^{ - 1}}\left( {{\boldsymbol{D}}_{\gamma ;k}^{{\mho _i}}{{\boldsymbol{C}}^{{\mho _i}}}} \right) \hfill \\ 
\end{gathered}  \right\}^{ - 1}}. \hfill \\ 
\end{gathered}  
\end{split}
\end{equation}
We then apply the following matrix inversion lemma \cite{chen2017maximum}
\begin{equation}
\begin{split}
\begin{gathered}
  {\left( {{\boldsymbol{G}} + {\boldsymbol{BCD}}} \right)^{ - 1}} \hfill \\
   = {{\boldsymbol{G}}^{ - 1}} - {{\boldsymbol{G}}^{ - 1}}{\boldsymbol{B}}{\left( {{{\boldsymbol{C}}^{ - 1}} + {\boldsymbol{D}}{{\boldsymbol{G}}^{ - 1}}{\boldsymbol{B}}} \right)^{ - 1}}{\boldsymbol{D}}{{\boldsymbol{G}}^{ - 1}} \hfill \\ 
\end{gathered}  
\end{split}
\end{equation}
with the identifications
\begin{equation}
\begin{split}
\left\{ \begin{gathered}
  {\boldsymbol{G}} = {\left[ {{{\left( {{\boldsymbol{B}}_{P\left( {k|k - 1} \right)}^i} \right)}^{ - 1}}} \right]^{\text{T}}}{\boldsymbol{\Lambda }}_{x;k}^i{\left( {{\boldsymbol{B}}_{P\left( {k|k - 1} \right)}^i} \right)^{ - 1}}, \hfill \\
  {\boldsymbol{B}} = {\left( {{\boldsymbol{D}}_{\gamma ;k}^{{\mho _i}}{{\boldsymbol{C}}^{{\mho _i}}}} \right)^{\text{T}}}, \hfill \\
  {\boldsymbol{C}} = {\left[ {{{\left( {{\boldsymbol{B}}_{R;k}^i} \right)}^{ - 1}}} \right]^{\text{T}}}{\boldsymbol{\Lambda }}_{y;k}^i{\left( {{\boldsymbol{B}}_{R;k}^i} \right)^{ - 1}}, \hfill \\
  {\boldsymbol{D}} = {\boldsymbol{D}}_{\gamma ;k}^{{\mho _i}}{{\boldsymbol{C}}^{{\mho _i}}}, \hfill \\ 
\end{gathered}  \right.
\end{split}
\end{equation}
and obtain the expression (\ref{equ:jt1jkpkib}) given below.
\newcounter{mytempeqncnt}
\begin{figure*}[!t]
\normalsize
\setcounter{mytempeqncnt}{\value{equation}}
\begin{equation}\label{equ:jt1jkpkib}
\begin{gathered}
\begin{gathered}
  {\left[ {{{\left( {{\boldsymbol{W}}_k^i} \right)}^{\text{T}}}{\boldsymbol{\Lambda }}_k^i{\boldsymbol{W}}_k^i} \right]^{ - 1}} = {\boldsymbol{B}}_{P\left( {k|k - 1} \right)}^i{\left( {{\boldsymbol{\Lambda }}_{x;k}^i} \right)^{ - 1}}{\left( {{\boldsymbol{B}}_{P\left( {k|k - 1} \right)}^i} \right)^{\text{T}}} - {\boldsymbol{B}}_{P\left( {k|k - 1} \right)}^i{\left( {{\boldsymbol{\Lambda }}_{x;k}^i} \right)^{ - 1}}{\left( {{\boldsymbol{B}}_{P\left( {k|k - 1} \right)}^i} \right)^{\text{T}}}{\left( {{\boldsymbol{D}}_{\gamma ;k}^{{\mho _i}}{C^{{\mho _i}}}} \right)^{\text{T}}} \hfill \\
  {\left[ {{\boldsymbol{B}}_{R;k}^i{{\left( {{\boldsymbol{\Lambda }}_{y;k}^i} \right)}^{ - 1}}{{\left( {{\boldsymbol{B}}_{R;k}^i} \right)}^{\text{T}}} + \left( {{\boldsymbol{D}}_{\gamma ;k}^{{\mho _i}}{{\boldsymbol{C}}^{{\mho _i}}}} \right){\boldsymbol{B}}_{P\left( {k|k - 1} \right)}^i{{\left( {{\boldsymbol{\Lambda }}_{x;k}^i} \right)}^{ - 1}}{{\left( {{\boldsymbol{B}}_{P\left( {k|k - 1} \right)}^i} \right)}^{\text{T}}}{{\left( {{\boldsymbol{D}}_{\gamma ;k}^{{\mho _i}}{C^{{\mho _i}}}} \right)}^{\text{T}}}} \right]^{ - 1}} \hfill \\
   \times \left( {{\boldsymbol{D}}_{\gamma ;k}^{{\mho _i}}{{\boldsymbol{C}}^{{\mho _i}}}} \right){\boldsymbol{B}}_{P\left( {k|k - 1} \right)}^i{\left( {{\boldsymbol{\Lambda }}_{x;k}^i} \right)^{ - 1}}{\left( {{\boldsymbol{B}}_{P\left( {k|k - 1} \right)}^i} \right)^{\text{T}}} \hfill \\ 
\end{gathered}  
\end{gathered} 
\end{equation}
\hrulefill
\vspace*{4pt}
\end{figure*}
According to (\ref{equ:jkiociuf}), (\ref{equ:kidkiatkwi}), and (\ref{equ:jkktjsikesg}), we obtain the following:
\begin{equation}\label{equ:kious1fkrikb}
\begin{split}
\begin{gathered}
  {\left( {{\boldsymbol{W}}_k^i} \right)^{\text{T}}}{\boldsymbol{\Lambda }}_k^i{\boldsymbol{D}}_k^i \hfill \\
   = {\left[ {{{\left( {{\boldsymbol{B}}_{P\left( {k|k - 1} \right)}^i} \right)}^{ - 1}}} \right]^{\text{T}}}{\boldsymbol{\Lambda }}_{x;k}^i{\left( {{\boldsymbol{B}}_{P\left( {k|k - 1} \right)}^i} \right)^{ - 1}}{\boldsymbol{\hat x}}_{k|k - 1}^i \hfill \\
   + {\left( {{\boldsymbol{D}}_{\gamma ;k}^{{\mho _i}}{{\boldsymbol{C}}^{{\mho _i}}}} \right)^{\text{T}}}{\left[ {{{\left( {{\boldsymbol{B}}_{R;k}^i} \right)}^{ - 1}}} \right]^{\text{T}}}{\boldsymbol{\Lambda }}_{y;k}^i{\left( {{\boldsymbol{B}}_{R;k}^i} \right)^{ - 1}}{\boldsymbol{s}}_k^{{\mho _i}}. \hfill \\ 
\end{gathered}
\end{split}
\end{equation}
Substituting formulas (\ref{equ:jt1jkpkib}) and (\ref{equ:kious1fkrikb}) into (\ref{equ:kidkiatkwi}) yields
\begin{equation}\label{equ:tijksxiug}
\begin{split}
{\boldsymbol{\hat x}}_{k|k}^i = {\boldsymbol{\hat x}}_{k|k - 1}^i + {\boldsymbol{\bar K}}_k^i\left[ {{\boldsymbol{s}}_k^{{\mho _i}} - {\boldsymbol{D}}_{\gamma ;k}^{{\mho _i}}{{\boldsymbol{C}}^{{\mho _i}}}{\boldsymbol{\hat x}}_{k|k - 1}^i} \right],
\end{split}
\end{equation}
where 
\begin{equation}
\begin{split}
\left\{ \begin{gathered}
  {\boldsymbol{\bar K}}_k^i = {\boldsymbol{\bar P}}_{k|k - 1}^i{\left( {{\boldsymbol{D}}_{\gamma ;k}^{{\mho _i}}{{\boldsymbol{C}}^{{\mho _i}}}} \right)^{\text{T}}} \hfill \\
  \left[ {{\boldsymbol{D}}_{\gamma ;k}^{{\mho _i}}{{\boldsymbol{C}}^{{\mho _i}}}{\boldsymbol{\bar P}}_{k|k - 1}^i{{\left( {{\boldsymbol{D}}_{\gamma ;k}^{{\mho _i}}{{\boldsymbol{C}}^{{\mho _i}}}} \right)}^{\text{T}}} + {\boldsymbol{\bar R}}_k^i} \right], \hfill \\
  {\boldsymbol{\bar P}}_{k|k - 1}^i = {\boldsymbol{B}}_{P\left( {k|k - 1} \right)}^i{\left( {{\boldsymbol{\Lambda }}_{x;k}^i} \right)^{ - 1}}{\left( {{\boldsymbol{B}}_{P\left( {k|k - 1} \right)}^i} \right)^{\text{T}}}, \hfill \\
  {\boldsymbol{\bar R}}_k^i = {\boldsymbol{B}}_{R;k}^i{\left( {{\boldsymbol{\Lambda }}_{y;k}^i} \right)^{ - 1}}{\left( {{\boldsymbol{B}}_{R;k}^i} \right)^{\text{T}}}. \hfill \\ 
\end{gathered}  \right.
\end{split}
\end{equation}
\textbf{Remark 1.} Equation (\ref{equ:tijksxiug}) is the optimal solution of ${{\boldsymbol{x}}_k^i}$, and it depends on the prior estimate ${{\boldsymbol{\hat x}}_{k|k - 1}^i}$ and available observation information ${{{\boldsymbol{s}}_k^{{\mho _i}}}}$. The observation information obtainable by the ${i}$th node is determined by the matrix ${{{\boldsymbol{D}}_{\gamma ;k}^{{\mho _i}}}}$.

According to the above derivations, we summarize the steps of the proposed DMCKF-DPD algorithm as follows.
\begin{enumerate}
\item Initialize the values of the ${\sigma }$, ${\varepsilon }$ (a small positive number), ${{\boldsymbol{\hat x}}_{0|0}^i}$, and ${{\boldsymbol{P}}_{0|0}^i}$.
\item Apply (\ref{equ:tkikbkikb}) to calculate ${{\boldsymbol{B}}_k^i}$, and calculate ${{\boldsymbol{\hat x}}_{k|k - 1}^i}$ and ${{\boldsymbol{P}}_{k|k - 1}^i}$ as follows:
\begin{equation}\label{equ:kixaki}
\begin{split}
\begin{gathered}
  {\boldsymbol{x}}_k^i = {\boldsymbol{Ax}}_{k - 1}^i, \hfill \\
  {\boldsymbol{P}}_{k|k - 1}^i = {\boldsymbol{AP}}_{k - 1|k - 1}^i{{\boldsymbol{A}}^{\text{T}}} + {\boldsymbol{Q}}. \hfill \\ 
\end{gathered} 
\end{split}
\end{equation}
\item Update the parameters ${{\boldsymbol{D}}_k^i = {\boldsymbol{W}}_k^i{\boldsymbol{x}}_k^i + {\boldsymbol{e}}_k^i}$ in the new system.
\item Utilize the fixed-point iterative algorithm to estimate the state of system as follows:
\begin{equation} \label{equ:kkiick}
\begin{split}
\begin{gathered}
  {\left( {{\boldsymbol{\hat x}}_{k|k}^i} \right)_{t + 1}} \hfill \\
   = {\boldsymbol{\hat x}}_{k|k - 1}^i + {\boldsymbol{\bar K}}_k^i\left[ {{\boldsymbol{s}}_k^{{\mho _i}} - {\boldsymbol{D}}_{\gamma ;k}^{{\mho _i}}{{\boldsymbol{C}}^{{\mho _i}}}{\boldsymbol{\hat x}}_{k|k - 1}^i} \right], \hfill \\ 
\end{gathered} 
\end{split}
\end{equation}
where 
\begin{equation} \label{equ:kirptikc}
\begin{split}
\begin{gathered}
  {\boldsymbol{\tilde K}}_k^i = {\boldsymbol{\tilde P}}_{k|k - 1}^i{\left( {{{\boldsymbol{C}}^{{\mho _i}}}} \right)^{\text{T}}}{\left( {{\boldsymbol{D}}_{\gamma ;k}^{{\mho _i}}} \right)^{\text{T}}} \times  \hfill \\
  \left[ {{\boldsymbol{D}}_{\gamma ;k}^{{\mho _i}}{{\boldsymbol{C}}^{{\mho _i}}}{\boldsymbol{\bar P}}_{k|k - 1}^i{{\left( {{{\boldsymbol{C}}^{{\mho _i}}}} \right)}^{\text{T}}}{{\left( {{\boldsymbol{D}}_{\gamma ;k}^{{\mho _i}}} \right)}^{\text{T}}} + {\boldsymbol{\tilde R}}_k^i} \right], \hfill \\ 
\end{gathered} 
\end{split}
\end{equation}
\begin{equation} \label{equ:trkibyi}
\begin{split}
{\boldsymbol{\tilde P}}_{k|k - 1}^i = {\boldsymbol{B}}_{P\left( {k|k - 1} \right)}^i{\left( {{\boldsymbol{\tilde \Lambda }}_{x;k}^i} \right)^{ - 1}}{\left( {{\boldsymbol{B}}_{P\left( {k|k - 1} \right)}^i} \right)^{\text{T}}},
\end{split}
\end{equation}
\begin{equation} \label{equ:tkrikyi}
\begin{split}
{\boldsymbol{\tilde R}}_k^i = {\boldsymbol{B}}_{R;k}^i{\left( {{\boldsymbol{\tilde \Lambda }}_{y;k}^i} \right)^{ - 1}}{\left( {{\boldsymbol{B}}_{R;k}^i} \right)^{\text{T}}},
\end{split}
\end{equation}
\begin{equation} \label{equ:kinegkie}
\begin{split}
{\boldsymbol{\tilde \Lambda }}_{x;k}^i = \operatorname{diag} \left[ {{\operatorname{G} _\sigma }\left( {\tilde e_k^{i;1}} \right), \cdots ,{\operatorname{G} _\sigma }\left( {\tilde e_k^{i;n}} \right)} \right],
\end{split}
\end{equation}
\begin{equation} \label{equ:kngegknig}
\begin{split}
{\boldsymbol{\tilde \Lambda }}_{y;k}^i = \operatorname{diag} \left[ {{\operatorname{G} _\sigma }\left( {\tilde e_k^{i;n + 1}} \right), \cdots ,{\operatorname{G} _\sigma }\left( {\tilde e_k^{i;n + {m_{i;a}}}} \right)} \right],
\end{split}
\end{equation}
\begin{equation} \label{equ:tkkikihw}
\begin{split}
\tilde e_k^{i;h} = d_k^{i;h} - {\boldsymbol{w}}_k^{i;h}{\left( {{\boldsymbol{\hat x}}_{k|k}^{i;h}} \right)_t}.
\end{split}
\end{equation}
\item Compare the state values obtained by the fixed-point iterative algorithm at ${t}$ and ${t+1}$. If these two state estimates satisfy the following criterion: 
\begin{equation} \label{equ:sgmxydys}
\begin{split}
\frac{{\left\| {{{\left( {{\boldsymbol{\hat x}}_{k|k}^i} \right)}_{t + 1}} - {{\left( {{\boldsymbol{\hat x}}_{k|k}^i} \right)}_t}} \right\|}}{{{{\left( {{\boldsymbol{\hat x}}_{k|k}^i} \right)}_t}}} \leqslant \varepsilon ,
\end{split}
\end{equation}
where ${\varepsilon }$ is the termination condition of the fixed-point iterative algorithm, set the optimal estimate of ${{\boldsymbol{\hat x}}_{k|k}^i}$ equal to ${{({\boldsymbol{\hat x}}_{k|k}^i)_{t + 1}}}$, and continue to step 6). Otherwise, set ${t \leftarrow t + 1}$ and return to step 4).
\item Update the covariance matrix as follows:
\begin{equation}\label{equ:dikikbkoikri}
\begin{split}
\begin{gathered}
  {\boldsymbol{P}}_{k|k}^i = \left( {{\boldsymbol{I}} - {\boldsymbol{\tilde K}}_k^i{\boldsymbol{D}}_{\gamma ;k}^{{\mho _i}}{{\boldsymbol{C}}^{{\mho _i}}}} \right){\boldsymbol{P}}_{k|k - 1}^i \hfill \\
  \left[ {{\boldsymbol{I}} - {\boldsymbol{\tilde K}}_k^i{{\left( {{{\boldsymbol{C}}^{{\mho _i}}}} \right)}^{\text{T}}}{{\left( {{\boldsymbol{D}}_{\gamma ;k}^{{\mho _i}}} \right)}^{\text{T}}}} \right] + {\boldsymbol{\tilde K}}_k^i{\boldsymbol{R}}_k^{{\mho _i}}{\left( {{\boldsymbol{\tilde K}}_k^i} \right)^{\text{T}}}, \hfill \\ 
\end{gathered}  
\end{split}
\end{equation}
set ${k \leftarrow k + 1}$, and return to step 2).
\end{enumerate}
\textbf{Remark 2.} 
The kernel bandwidth ${\sigma }$ is an important parameter in the MC criterion. Broadly speaking, the convergence rate of the algorithm increases with increasing ${\sigma }$, but the accuracy of the algorithm decreases with increasing ${\sigma }$. 

\subsection{Computational complexity}
The computational complexity of the proposed DMCKF-DPD algorithm can be compared with that of the stationary DKF \cite{8409298} in terms of the equations and operations employed by the two algorithms, which are listed in Table 1.

The stationary DKF algorithm mainly includes (\ref{equ:aa}), (\ref{equ:kiyikd}), and (\ref{equ:kixaki}) cited in the present paper, and (5) and (6) cited within\cite{8409298}. Therefore, we can evaluate the computational complexity of the stationary DKF as
\begin{equation}\label{equ:iugoiug}
\begin{split}
\begin{gathered}
  {S_{SDKF}} = 11{n^3} + 12m_{i;a}^2n + 10{m_{i;a}}{n^2} + 4m_{i;a}^2 \hfill \\
   - 2{m_{i;a}}n - {n^2} - 2n - {m_{i;a}} + 2O(m_{i;a}^3). \hfill \\ 
\end{gathered}  
\end{split}
\end{equation}
The proposed algorithm mainly includes (\ref{equ:aa}) and (\ref{equ:kiyikd}), (\ref{equ:kixaki}), (\ref{equ:epxeeg}), and (\ref{equ:kkiick})-(\ref{equ:tkkikihw}). Accordingly, the computational complexity of the DMCKF-DPD algorithm can be defined based on an average number of fixed-point iterative algorithm iterations ${T}$ as 
\begin{equation}\label{equ:ijgotno}
\begin{split}
\begin{gathered}
  {S_{{\text{SDMCKF}}}} \hfill \\
  {\text{ = 6}}Tm_{i;a}^3 + 6T{n^3} + {\text{16}}Tm_{i;a}^2n + 10T{m_{i;a}}{n^2} \hfill \\
   + (2 - 3T)m_{i;a}^2 + 2{n^2} + (2 - 3T){m_{i;a}}n +  \hfill \\
  ({\text{6}}T - 1){m_{i;a}} + ({\text{6}}T - 1)n + T{\text{O(}}{n^3}{\text{)}} + T{\text{O(}}m_{i;a}^3{\text{)}}. \hfill \\ 
\end{gathered} 
\end{split}
\end{equation}

\begin{table}\label{table1}
\caption{Computational complexities of the  proposed DMCKF-DPD algorithm and the stationary DKF algorithm\cite{8409298}}  
\begin{tabular*}{8.9cm}{lll}  
\hline  
Equation & Addition/subtraction     & Division,\\
         & and multiplication       & matrix inversion,\\ 
         &                          & Cholesky \\
         &                          & decomposition,\\
         &                          & and exponentiation\\
\hline  
(\ref{equ:aa}) & ${2n{m_{i;a}}}$                                 & 0\\
(\ref{equ:kiyikd}) & ${2m_{i;a}^2 - {m_{i;a}}}$            & 0\\
(\ref{equ:kixaki})   & ${2{n^2} - n}$                                  & 0\\
{(5)} in\cite{8409298}  & ${\begin{gathered}
  9{n^3} - 4{n^2} + 6{m_{i;a}}{n^2} +  \hfill \\
  4m_{i;a}^2n - 3{m_{i;a}}n + m_{i;a}^2 \hfill \\ 
\end{gathered} }$                                                                                      & ${{\text{O(}}m_{i;a}^3{\text{)}}}$\\
{(6a)} in\cite{8409298} & ${{\text{2}}m_{i;a}^2n + 3{m_{i;a}}n - n + 2{n^2}}$               & 0\\ 
{(6b)} in\cite{8409298} & ${\begin{gathered}
  {\text{2}}{n^3} - {n^2} + {\text{4}}{m_{i;a}}{n^2} +  \hfill \\
  {\text{6}}m_{i;a}^2n - {\text{4}}{m_{i;a}}n + m_{i;a}^2{\text{ }} \hfill \\ 
\end{gathered} }$                                                                                      & ${{\text{O(}}m_{i;a}^3{\text{)}}}$\\
(\ref{equ:epxeeg})   & ${2n + 4}$                                                                      & 2\\
(\ref{equ:kkiick})   & ${2m_{i;a}^2n + {\text{3}}{m_{i;a}}n}$                                          & 0\\
(\ref{equ:kirptikc})   & ${\begin{gathered}
  2m_{i;a}^3 + 8m_{i;a}^2n + 4{m_{i;a}}{n^2} -  \hfill \\
  5{m_{i;a}}n - m_{i;a}^2 \hfill \\ 
\end{gathered} }$                                                                       			  & 0\\
(\ref{equ:trkibyi})   & ${4{n^3} + {\text{4}}n}$                                        			  & ${2n + {\text{O}}({n^3}){\text{ }}}$\\
(\ref{equ:tkrikyi})   & ${2{m_{i;a}}n + 4{m_{i;a}} + 4m_{i;a}^3 - 2m_{i;a}^2}$           			  & ${2{m_{i;a}} + {\text{O}}(m_{i;a}^3)}$\\
(\ref{equ:kinegkie})   & ${2{n^2} + {\text{4}}n}$                                                      & ${2n}$\\  
(\ref{equ:kngegknig})   & ${2{m_{i;a}}n + {\text{4}}{m_{i;a}}}$                            		  & ${2{m_{i;a}}}$\\    
(\ref{equ:tkkikihw})   & ${2n{\text{ }}}$                                                 		       & 0\\   
\hline  
\end{tabular*}  
\end{table} 

We can infer from this discussion that the computational complexity of the DMCKF-DPD algorithm is moderate compared with that of the stationary DKF, provided that the value of ${T}$ is small, which is indeed the case, as will be demonstrated later in Section \ref{section:simulations}. 
\subsection{Convergence issue}
A full analysis of the convergence behavior of the DMCKF-DPD algorithm based on the fixed-point iterative algorithm is very complicated. Therefore, we give only a sufficient condition that ensures the convergence of the fixed-point iterative algorithm.  However, a detailed proof process is not presented here because the convergence condition is similar to the analysis presented in an earlier work\cite{chen2017maximum}, which can be consulted for additional details.

\textbf{Theorem 1.} First, we assume the conditions ${{\beta _i} > {\zeta _i} = \frac{{\sqrt n \sum\limits_{h = 1}^L {{{\left\| {{{\left( {{\boldsymbol{w}}_k^{i;h}} \right)}^{\text{T}}}} \right\|}_1}\left| {d_k^{i;h}} \right|} }}{{{\lambda _{\min }}\sum\limits_{h = 1}^L {{{\left( {{\boldsymbol{w}}_k^{i;h}} \right)}^{\text{T}}}{\boldsymbol{w}}_k^{i;h}} }}}$ and ${{\sigma _i} \ge \max {\rm{\{ }}\sigma _i^*,\sigma _i^\diamondsuit {\rm{\} }}}$. Here, ${\sigma _i^*}$ is the optimal result of the equation ${\phi ({\sigma _i}) = {\beta _i}}$, where 
\newcounter{mytempeqncnt1}
\begin{figure*}[!t]
\normalsize
\setcounter{mytempeqncnt1}{\value{equation}}
\begin{equation}\label{equ:wlindsisikc}
\psi \left( {{\sigma _i}} \right) = \frac{{\sqrt n \sum\limits_{h = 1}^L {\left[ {\left( {{\beta _i}{{\left\| {{\boldsymbol{w}}_k^{i;h}} \right\|}_1} + \left| {d_k^{i;h}} \right|} \right){{\left\| {{\boldsymbol{w}}_k^{i;h}} \right\|}_1} \times \left( {{\beta _i}{{\left\| {{{\left( {{\boldsymbol{w}}_k^{i;h}} \right)}^{\text{T}}}{\boldsymbol{w}}_k^{i;h}} \right\|}_1} + {{\left\| {{{\left( {{\boldsymbol{w}}_k^{i;h}} \right)}^{\text{T}}}d_k^{i;h}} \right\|}_1}} \right)} \right]} }}{{\sigma _i^2{\lambda _{\min }}\left[ {\sum\limits_{h = 1}^L {{\operatorname{G} _{{\sigma _i}}}\left( {{\beta _i}{{\left\| {{\boldsymbol{w}}_k^{i;h}} \right\|}_1} + \left| {d_k^{i;h}} \right|} \right){{\left( {{\boldsymbol{w}}_k^{i;h}} \right)}^{\text{T}}}{\boldsymbol{w}}_k^{i;h}} } \right]}},{\sigma _i} \in (0,\infty ) 
\end{equation}
\vspace*{4pt}
\end{figure*}
\begin{equation}\label{equ:kddokz}
\begin{split}
\begin{gathered}
  \phi \left( {{\sigma _i}} \right) = \frac{{\sqrt n \sum\limits_{h = 1}^L {{{\left\| {{{\left( {{\boldsymbol{w}}_k^{i;h}} \right)}^{\text{T}}}} \right\|}_1}\left| {d_k^{i;h}} \right|} }}{{{\lambda _{\min }}\left[ {\sum\limits_{h = 1}^L {{\operatorname{G} _{{\sigma _i}}}\left( {{\beta _i}{{\left\| {{\boldsymbol{w}}_k^{i;h}} \right\|}_1} + \left| {d_k^{i;h}} \right|} \right){{\left( {{\boldsymbol{w}}_k^{i;h}} \right)}^{\text{T}}}{\boldsymbol{w}}_k^{i;h}} } \right]}}, \hfill \\
  {\sigma _i} \in \left( {0,\infty } \right), \hfill \\ 
\end{gathered} 
\end{split}
\end{equation}
and ${\sigma _i^\diamondsuit }$ is the result of the equation ${\psi ({\sigma _i}) = {\alpha _i}(0 < {\alpha _i} < 1)}$, where  ${\psi ({\sigma _i})}$is given in (\ref{equ:wlindsisikc}) below. Accordingly, it holds that ${||\operatorname{f} ({\boldsymbol{x}}_k^i)|{|_1} \leqslant {\beta _i}}$ and ${||{\nabla _{{\boldsymbol{x}}_k^i}}\operatorname{f} ({\boldsymbol{x}}_k^i)|{|_1} \leqslant {\alpha _i}}$ for all ${{\boldsymbol{x}}_k^i \in \{ {\boldsymbol{x}}_k^i \in {\mathbb{R}^n}:||{\boldsymbol{x}}_k^i|{|_1} \leqslant {\beta _i}\} }$. Here, the ${n \times n}$ Jacobian matrix ${\operatorname{f} ({\boldsymbol{x}}_k^i)}$ is given as follows:
\begin{equation}
\begin{split}
{\nabla _{{\boldsymbol{x}}_k^i}}\operatorname{f} \left( {{\boldsymbol{x}}_k^i} \right) = \left[ {\frac{\partial }{{\partial {\boldsymbol{x}}_k^{i;1}}}\operatorname{f} \left( {{\boldsymbol{x}}_k^i} \right) \cdots \frac{\partial }{{\partial {\boldsymbol{x}}_k^{i;n}}}\operatorname{f} \left( {{\boldsymbol{x}}_k^i} \right)} \right],
\end{split}
\end{equation} 
where the terms are defined in (\ref{equ:kddokzkkhiw}) below
\newcounter{mytempeqncnt2}
\begin{figure*}[!t]
\normalsize
\setcounter{mytempeqncnt2}{\value{equation}}
\begin{equation}\label{equ:kddokzkkhiw}
\frac{\partial }{{\partial {\boldsymbol{x}}_k^{i;j}}}\operatorname{f} \left( {{\boldsymbol{x}}_k^i} \right) =  - {{\boldsymbol{T}}_g}\frac{1}{{{\sigma ^2}}}\sum\limits_{i = 1}^L {\left[ {e_k^{i;h}w_k^{i;h;j}{\operatorname{G} _\sigma }\left( {e_k^i} \right){{\left( {{\boldsymbol{w}}_k^{i;h}} \right)}^{\text{T}}}{\boldsymbol{w}}_k^{i;h}} \right]} \operatorname{f} \left( {{\boldsymbol{x}}_k^i} \right) + {{\boldsymbol{T}}_g}\frac{1}{{{\sigma ^2}}}\sum\limits_{i = 1}^L {\left[ {e_k^{i;h}w_k^{i;h;j}{\operatorname{G} _\sigma }\left( {e_k^i} \right){{\left( {{\boldsymbol{w}}_k^{i;h}} \right)}^{\text{T}}}d_k^{i;h}} \right]}
\end{equation}
\hrulefill
\vspace*{4pt}
\end{figure*}
with 
\begin{equation}
\begin{split}
{{\boldsymbol{T}}_g} = {\left[ {\sum\limits_{h = 1}^L {{\operatorname{G} _\sigma }\left( {d_k^{i;h} - {\boldsymbol{w}}_k^{i;h}{\boldsymbol{x}}_k^i} \right){{\left( {{\boldsymbol{w}}_k^{i;h}} \right)}^{\text{T}}}{\boldsymbol{w}}_k^{i;h}} } \right]^{ - 1}},
\end{split}
\end{equation} 
and ${{w_k^{i;h;j}}}$ is the ${j}$th element of the vector ${{{\boldsymbol{w}}_k^{i;h}}}$.

According to Theorem 1, we obtain the following conditions: 
\begin{equation}
\begin{split}
\left\{ {\begin{array}{*{20}{l}}
  {{{\left\| {\operatorname{f} \left( {{\boldsymbol{x}}_k^i} \right)} \right\|}_1} \leqslant \beta ,} \\ 
  {{{\left\| {{\nabla _{x_k^i}}\operatorname{f} \left( {{\boldsymbol{x}}_k^i} \right)} \right\|}_1} \leqslant \alpha  < 1.} 
\end{array}} \right.
\end{split}
\end{equation} 
if the kernel bandwidth ${\sigma }$ is sufficiently large (e.g., greater than ${\max {\text{\{ }}\sigma _i^*,\sigma _i^\diamondsuit {\text{\} }}}$). According to the Banach fixed-point theorem the fixed-point iterative algorithm in DMCKF-DPD will surely converge to a unique fixed point in the range ${{\boldsymbol{x}}_k^i \in \{ {\boldsymbol{x}}_k^i \in {\mathbb{R}^n}: ||{\boldsymbol{x}}_k^i|{|_1} \leqslant {\beta _i}\} }$ provided that the initial state of the system meets the condition ${||{({\boldsymbol{x}}_k^i)_0}|| \leqslant \beta }$ and ${\sigma }$ is sufficiently large.

Theorem 1 demonstrates that the kernel bandwidth of the Gaussian kernel function has a important influence on the convergence of the DMCKF-DPD algorithm. Here, reducing the kernel bandwidth can improve the accuracy of state estimation, but this will also decrease the convergence rate of the algorithm or make it diverge. Conversely, increasing the kernel width will increase the convergence rate of the algorithm, but will often yield poor estimation performance under impulsive noise conditions. In practice, the kernel bandwidth can be selected by trial and error in accordance with the desired estimation accuracy and convergence rate of the algorithm.

\section{simulations}\label{section:simulations}
We compared the state estimation performances obtained by the proposed DMCKF-DPD algorithm and the stationary DKF algorithm\cite{8409298} when applying them to a classic WSN example \cite{5411741} under data packet drop and impulsive noise conditions. The network includes 20 sensor nodes (${N=20}$), and the nodes and connections are illustrated in Fig. 1. We consider the state estimation problem with the following state and observation equations:
\begin{equation}
\begin{split}
\left[ {\begin{array}{*{20}{c}}
  {x_{k + 1}^{i;1}} \\ 
  {x_{k + 1}^{i;2}} \\ 
  {x_{k + 1}^{i;3}} 
\end{array}} \right] = \left[ {\begin{array}{*{20}{c}}
  1&{\Delta T}&{\Delta {T^2}/2} \\ 
  0&1&{\Delta T} \\ 
  0&0&1 
\end{array}} \right]\left[ {\begin{array}{*{20}{c}}
  {x_k^{i;1}} \\ 
  {x_k^{i;2}} \\ 
  {x_k^{i;3}} 
\end{array}} \right] + {\boldsymbol{q}}_k^i,
\end{split}
\end{equation} 
\begin{equation}
\begin{split}
{\boldsymbol{y}}_k^i = \left[ {\begin{array}{*{20}{c}}
  0&1&0 
\end{array}} \right]\left[ {\begin{array}{*{20}{c}}
  {x_k^{i;1}} \\ 
  {x_k^{i;2}} \\ 
  {x_k^{i;3}} 
\end{array}} \right] + {\boldsymbol{v}}_k^i.
\end{split}
\end{equation}
Here, ${\Delta T{\text{ = 0}}{\text{.1}}}$ s represents the measurement time interval, ${{\boldsymbol{x}}_k^i = {[x_k^{i;1}{\text{  }}x_k^{i;2}{\text{  }}x_k^{i;3}]^{\text{T}}}}$ represent the position, velocity, and acceleration of the target, respectively, and ${{\boldsymbol{q}}_k^i}$ and ${{\boldsymbol{v}}_k^i}$ respectively represent mutually uncorrelated process noise and measurement noise (impulsive noise) with the following distributions:
\begin{equation}
\begin{split}
\begin{gathered}
  {\boldsymbol{q}}_k^{i;1} \sim 0.9N(0,{\text{ }}0.01) + 0.1N(0,{\text{ }}1), \hfill \\
  {\boldsymbol{q}}_k^{i;2} \sim 0.9N(0,{\text{ }}0.01) + 0.1N(0,{\text{ }}1), \hfill \\
  {\boldsymbol{q}}_k^{i;3} \sim 0.9N(0,{\text{ }}0.01) + 0.1N(0,{\text{ }}1), \hfill \\
  {{\boldsymbol{v}}_k} \sim 0.9N(0,{\text{ }}0.01) + 0.1N(0,{\text{ }}100). \hfill \\ 
\end{gathered}   
\end{split}
\end{equation}
In the simulations, we assume that only the velocity of the target can be observed. In addition, we set ${\varepsilon  = {10^{ - 6}}}$, and the initial values of the true state (state of the target), the estimated state, and the covariance matrix are respectively given as follows:
\begin{equation}
\begin{split}
\begin{gathered}
  {\boldsymbol{x}}_0^i = {\left[ {\begin{array}{*{20}{c}}
  0&0&1 
\end{array}} \right]^{\text{T}}}, \hfill \\
  {\boldsymbol{\hat x}}_{0|0}^i = {\left[ {\begin{array}{*{20}{c}}
  0&0&1 
\end{array}} \right]^{\text{T}}} + N(0,{\text{  0}}{\text{.01}}) \times {\left[ {\begin{array}{*{20}{c}}
  1&1&1 
\end{array}} \right]^{\text{T}}}, \hfill \\
  {\boldsymbol{P}}_{0|0}^i = 0.01 \times \operatorname{diag} \left\{ {\begin{array}{*{20}{c}}
  1&1&1 
\end{array}} \right\}. \hfill \\ 
\end{gathered}   
\end{split}
\end{equation}
The state estimation performance was measured using the mean square deviation (MSD), which can be defined for node ${i}$ at instant ${k}$ as follows:
\begin{equation}
\begin{split}
{\text{MSD}}_k^i: = E\left\{ {{{\left\| {x_k^i - \hat x_k^i} \right\|}^2}} \right\}. 
\end{split}
\end{equation}
The results listed in the tables are the averages of 100 independent trials, and each trial consists of 1000 iterations. 

\begin{figure}
\label{Fig1}
\centerline{\includegraphics[width=\columnwidth]{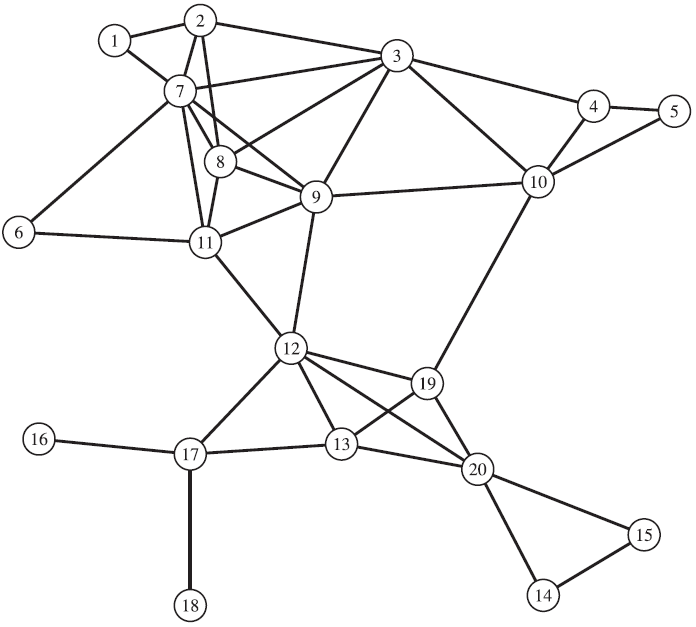}}
\caption{Network topology of the example WSN with ${N = 20}$ nodes}
\end{figure}

\begin{table}
\centering
\caption{Average number of iterations for every time step of the proposed DMCKF-DPD algorithm with different values of ${\sigma }$}  
\begin{tabular*}{8.0cm}{lccc} 
\hline  
 Average number   & ${p_k^{i;j} = 0.9}$  & ${p_k^{i;j} = 0.8}$     &${p_k^{i;j} = 0.7}$    \\
 of iterations                                                                                              \\
\hline
 ${\sigma  = 0.4}$   & 3.4760         & 3.9280      & 4.2500 \\
 ${\sigma  = 0.6}$   & 2.3850         & 2.6710      & 2.9460 \\
 ${\sigma  = 1.0}$   & 1.8620         & 1.9650      & 2.1730 \\
 ${\sigma  = 4.0}$   & 1.1640         & 1.2140      & 1.2480 \\
 ${\sigma  = 8.0}$   & 1.0690         & 1.0870      & 1.1020 \\
\hline
\end{tabular*}  
\end{table}

\begin{table*}
\centering
\caption{Average MSD values obtained under impulsive noise conditions for the distributed network}  
\begin{tabular*}{13.5cm}{lccccccc}
\hline   
 Node                                           & 16      & 5       & 4       & 2       & 8       & 9       & 7       \\                                                                                                
 \hline    
Number of neighbors                             & 1       & 2       & 3       & 4       & 5       & 6       & 7       \\
Stationary DKF  algorithm (${\sigma  = 2}$)                            & -2.1291 & -2.2715 & -2.2672 & -2.2812 & -2.2907 & -2.2857 & -2.3007 \\
DMCKF-DPD algorithm (${\sigma  = 2}$)        & -3.8419 & -3.9370 & -4.0695 & -4.0463 & -4.0989 & -4.2063 & -4.1700 \\
DMCKF-DPD algorithm (${\sigma  = 5}$)        & -3.8033 & -3.9126 & -4.0546 & -4.0387 & -4.0596 & -4.0722 & -4.1064 \\
DMCKF-DPD algorithm (${\sigma  = 10}$)       & -3.7800 & -3.9042 & -4.0376 & -4.0358 & -4.0417 & -4.0331 & -4.0926 \\
\hline
\end{tabular*}  
\end{table*}

\begin{figure}
\label{fig2}
\centerline{\includegraphics[width=\columnwidth]{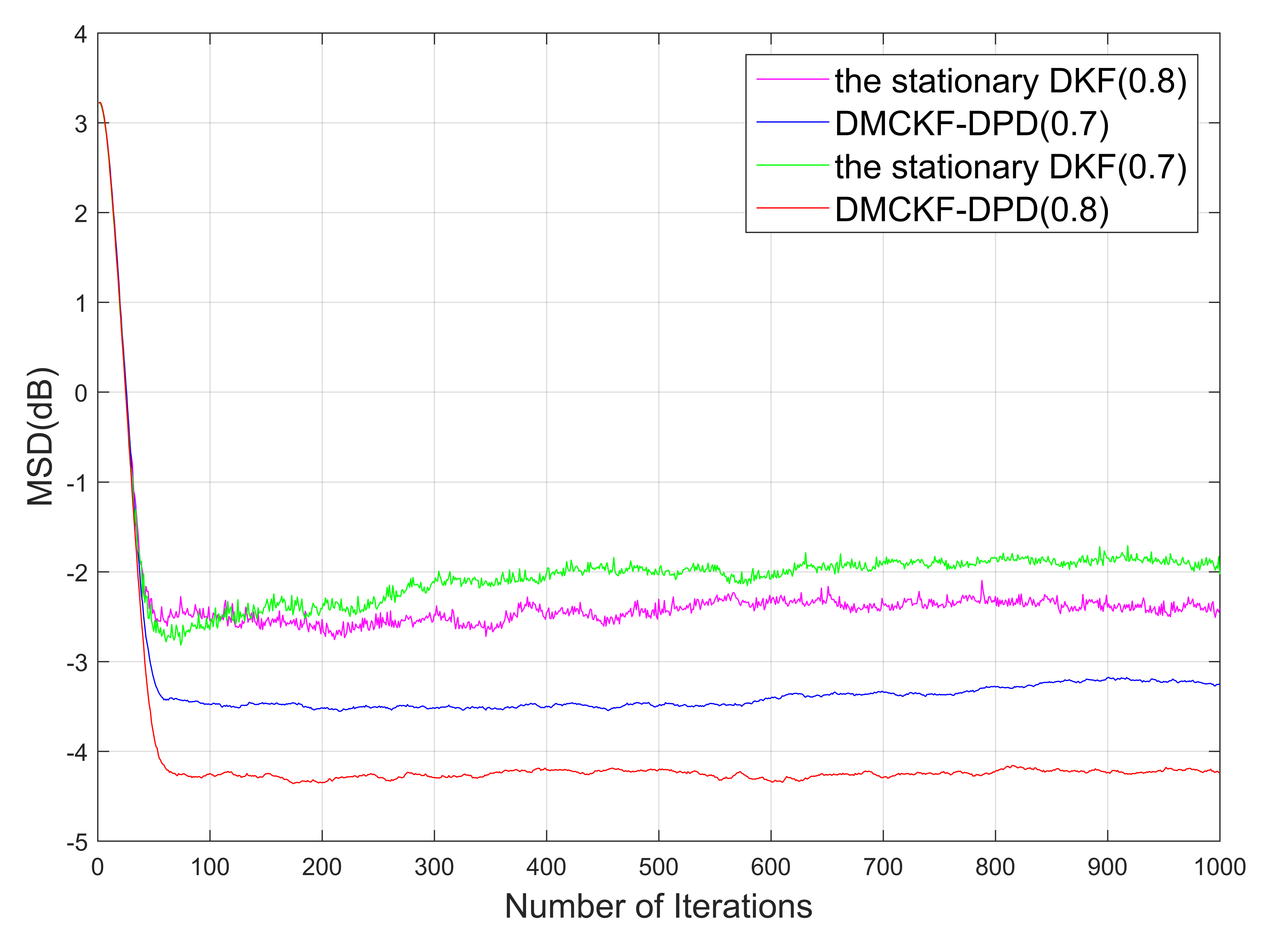}}
\caption{MSD values obtained by the state estimation algorithms under data packet drop and impulsive noise conditions for the distributed network (${\sigma  = 2}$).}
\end{figure}
The MSD values obtained by the two different algorithms with ${\sigma  = 2}$ are presented in Fig. 2 as a function of the number of iterations. Here, the labels Stationary DKF (0.8) and Stationary DKF (0.7) correspond to the stationary DKF algorithm with parameters ${p_k^{i;j} = 0.8}$ and ${p_k^{i;j} = 0.7}$, respectively. Similarly, the labels DMCKF-DPD (0.8) and DMCKF-DPD (0.7) correspond to the proposed algorithm with parameters ${p_k^{i;j} = 0.8}$ and ${p_k^{i;j} = 0.7}$, respectively. It is obvious that the DMCKF-DPD algorithm outperformed the conventional stationary DKF algorithm by about 1.5 dB when considering data packet drops and impulsive noise.

The average number of iterations required by the proposed DMCKF-DPD algorithm for convergence to an optimal solution at every time step are listed in Table 2 for different values of ${\sigma }$. We note that the convergence to the optimal solution is very fast because the initial value of the fixed-point iterative algorithm is set as (${{({\boldsymbol{\hat x}}_{k|k}^i)_0} = {\boldsymbol{\hat x}}_{k|k - 1}^i}$). It is also obvious that the number of iterations required by the fixed-point iterative algorithm decreases with increasing ${\sigma }$, and the convergence rate of the DMCKF-DPD algorithm correspondingly increases.

The average MSD values obtained at select nodes of the distributed system by the DMCKF-DPD and stationary DKF algorithms under impulsive noise conditions (${p_k^{i;j} = 0.8}$) are listed in Table 3. The proposed algorithm demonstrates better state estimation performance than the stationary DKF algorithm at all kernel bandwidths considered. We also note that the performance of the proposed algorithm is increasingly superior to that of the conventional algorithm as the number of neighbors for a given node in the network increases.

\section{Conclusion}\label{section:Conclusion}
This paper presented a novel distributed Kalman filter algorithm for state estimation in WSNs that accounts for data packet drops under non-Gaussian noise conditions. The approach requires each node in the WSN to recursively compute local state estimates in a collaborative manner with its neighboring nodes. The computational complexity of the DMCKF-DPD algorithm was demonstrated to be moderate compared to that of the conventional stationary DKF. In addition, a sufficient condition to ensure the convergence of the fixed-point iterative algorithm was presented. Simulations conducted with a 20-node WSN clearly demonstrated that the DMCKF-DPD algorithm provides significantly more accurate state estimation performance than the stationary DKF, particularly under non-Gaussian noise conditions. As future work, we intend to investigate the state estimation performance of a distributed Kalman filter based on the minimum error entropy (MEE) criterion for WSNs under data packet drop conditions.
\appendix

\bibliographystyle{unsrt}
\bibliography{ref}
\end{document}